\documentclass[10pt,DIV=16,twocolumn,numbers=noenddot]{scrartcl}

\usepackage[utf8]{inputenc}
\usepackage[T1]{fontenc}
\usepackage{lmodern}
\usepackage{amsmath,amssymb}
\usepackage[german]{authblk}
\usepackage[format=plain,labelfont={bf}]{caption}
\usepackage{listings}
\usepackage[shortcuts]{extdash} 
\usepackage{xcolor}
\usepackage{booktabs}
\usepackage{hyperref}
\usepackage[sorting=none,citestyle=numeric-comp,eprint=false,isbn=false]{biblatex}
\setkomafont{disposition}{\bfseries}
\allowdisplaybreaks

\newcommand{\mytitle}{An algorithm to approximate the real trilogarithm for a real argument}
\newcommand{\myauthor}{Alexander Voigt}
\newcommand{\id}{\operatorname{id}}
\newcommand{\Li}[1]{\operatorname{Li}_{#1}}
\newcommand{\li}[1]{\operatorname{li}_{#1}}
\newcommand{\appref}[1]{Appendix~\ref{#1}}

\newcommand{\tabref}[1]{\tablename~\ref{#1}}
\newcommand{\complexes}{\mathbb{C}}
\newcommand{\reals}{\mathbb{R}}
\newcommand{\dt}{\textnormal{d}t}

\hypersetup{
  pdftitle   = {\mytitle},
  pdfauthor  = {\myauthor},
  colorlinks = true,
  linkcolor  = blue,
  citecolor  = blue,
  urlcolor   = blue,
  linkcolor  = blue
}

\title{\mytitle}
\author{\myauthor}
\affil{Flensburg University of Applied Sciences,\\ Kanzleistra{\ss}e 91--93, 24943 Flensburg, Germany}
\date{\today}

\begin{document}
\maketitle

\section*{Abstract}

We present an algorithm to approximate the real trilogarithm for a
real argument with IEEE 754\-/1985 double precision accuracy. The
approximation is structured such that it can make use of
instruction\-/level parallelism when executed on appropriate CPUs.

\section{Introduction}

The trilogarithm $\Li{3}$ is a special function that appears in the
calculation of loop integrals in high\-/energy physics, for example in
two\-/loop self\-/energy integrals in the calculation of the Higgs
boson pole mass or in two\-/loop three\-/point integrals in the
calculation of Higgs boson decays. For this reason the trilogarithm is
implemented in Feynman integral libraries such as TSIL
\cite{Martin:2005qm} and 3vil \cite{Martin:2016bgz} or in the CHAPLIN
\cite{Buehler:2011ev}, HPOLY.f \cite{Ablinger:2018sat} and handyG
\cite{Naterop:2019xaf} libraries as one basis function to numerically
evaluate harmonic or generalized polylogarithms, respectively, in
terms of which certain classes of Feynman integrals can be
expressed. Since such loop integrals must be numerically evaluated
many times in parameter studies of models beyond the Standard Model of
particle physics, a time\-/efficient implementation of the
trilogarithm is advantageous.

An often used strategy to implement a special function is to map its
argument to one or more small regions on which a suitable approximant
for the function can be given. This strategy is, for example, usually
used to implement the real dilogarithm of a real argument: Using the
known functional relations for the real dilogarithm, it is possible to
map its argument to the interval $[0,1/2]$, where an expansion in
terms of Chebyshev polynomials \cite{luke} or a rational minimax
approximant \cite{morris,Voigt:2022xnc} can be used for the numerical
evaluation.
For a particularly time\-/efficient evaluation of a special function
on some interval one should try to minimize the number of costly
arithmetic floating\-/point operations, such as division or
multiplication. In addition one can try to make use of so\-/called
instruction\-/level parallelism (ILP), which is the ability of modern
CPUs to execute multiple independent operations at the same time. The
combination of all these strategies was for example used in
Ref.~\cite{Voigt:2022xnc} to obtain a time\-/efficient implementation
of the real dilogarithm for a real argument with IEEE 754\-/1985
double precision accuracy.  In this publication we will use the
strategy from Ref.~\cite{Voigt:2022xnc} to construct a
time\-/efficient algorithm for the numerical evaluation of the real
trilogarithm for a real argument with IEEE 754\-/1985 double precision
accuracy.

\section{The trilogarithm}
\label{sec:trilogaithm}

For all $z\in\complexes$ the trilogarithm
$\Li{3}:\complexes\to\complexes$ is defined as (see e.g.\
\cite{lewin})
\begin{equation}
  \label{eq:trilog}
  \Li{3}(z) = \int_0^z \frac{\Li{2}(t)}{t}\dt,
\end{equation}
where $\Li{2}:\complexes\to\complexes$ is the dilogarithm, defined as
\begin{equation}
  \label{eq:dilog}
  \Li{2}(z) = -\int_0^z \frac{\ln(1-t)}{t}\dt.
\end{equation}
For $|z|<1$ the trilogarithm has the series expansion
\begin{equation}
  \Li{3}(z) = \sum_{k=1}^\infty \frac{z^k}{k^3}.
  \label{eq:series}
\end{equation}
For $z\neq 0$ the following relations hold:
\begin{align}
  \Li{3}(z) &= \Li{3}(1/z) - \ln(-z)\zeta(2) - \frac{1}{6}\ln^3(-z), \label{eq:inversion1} \\
  \Li{3}(z) &= -\Li{3}(1-1/z) - \Li{3}(1-z) + \zeta(3) \nonumber \\
         &+ \ln(z)\zeta(2) - \frac{1}{2}\ln^2(z)\ln(1-z) + \frac{1}{6}\ln^3(z), \label{eq:inversion2}
\end{align}
where $\zeta$ is the Riemann zeta function with $\zeta(2)=\pi^2/6$.
For the following considerations we define the real trilogarithm for a
real argument, $\li{3}:\reals\to\reals$, as
\begin{equation}
  \label{eq:real_trilog}
  \li{3}(x) = \Re[\Li{3}(x)].
\end{equation}

\section{Algorithm to approximate the real trilogarithm}
\label{sec:algorithm}

To obtain a time\-/efficient approximation of the real trilogarithm we
proceed similarly to Ref.~\cite{Naterop:2019xaf}: We use
Eqs.~\eqref{eq:inversion1}--\eqref{eq:inversion2} to transform the
argument of the trilogarithm to the intervals $[-1,0]$ and/or
$[0,1/2]$.  On each of these intervals Ref.~\cite{Naterop:2019xaf}
approximates $\li{3}$ in terms of Chebyshev polynomials, which are
evaluated using Clenshaw's algorithm \cite{clenshaw}. Clenshaw's
algorithm, however, is purely sequential and thus cannot make use of
ILP. For this reason we use a different technique to approximate
$\li{3}$ on these intervals to allow for ILP: We use a rational
minimax approximant of the form
\begin{equation}
  \li{3}(x) =
  x\sum_{k=0}^\infty \frac{x^{k}}{(k+1)^3} \approx
  x\frac{\sum_{k=0}^5 p_k x^k}{\sum_{k=0}^6 q_k x^k},
  \label{eq:minimax}
\end{equation}
where we evaluate the numerator and denominator polynomials using
Estrin's scheme \cite{estrin}.  The coefficients $p_k$ and $q_k$ in
Eq.~\eqref{eq:minimax} are calculated using the
\texttt{MiniMaxApproximation} function from Wolfram/Mathematica
\cite{mathematica} and are listed in
\tabref{tab:coeffs_neg} and \ref{tab:coeffs_pos}. For arguments
$x\in[-1,0]$ the error of the approximant in Eq.~\eqref{eq:minimax} is
less than $2.050\cdot 10^{-17}$, while for $x\in[0,1/2]$ it is less
than $1.066\cdot 10^{-17}$. These maximum errors are small enough to
achieve IEEE 754\-/1985 double precision accuracy in the numeric
evaluation of $\li{3}$.

In detail, our algorithm to numerically evaluate $\li{3}$ is as
follows: We split the domain of $\li{3}$ into the sub\-/domains
$\reals=(-\infty,-1)\cup\{-1\}\cup(-1,0)\cup\{0\}\cup(0,1/2)\cup\{1/2\}\cup(1/2,1)\cup\{1\}\cup(1,2)\cup\{2\}\cup(2,\infty)$.
For arguments $x\in(-\infty,-1)$ we use Eq.~\eqref{eq:inversion1} to
transform the argument to the interval $(-1,0)$, where we use the
rational minimax approximant from Eq.~\eqref{eq:minimax} with the
coefficients listed in \tabref{tab:coeffs_neg}.  For $x=-1$ we
implement the known value $\li{3}(-1)=-3\zeta(3)/4$. For $x\in(-1,0)$
we directly use the appropriate approximant from
Eq.~\eqref{eq:minimax}. For $x=0$ we use the known value
$\li{3}(0)=0$. For $x\in(0,1/2)$ we directly use the approximant in
Eq.~\eqref{eq:minimax} with the coefficients listed in
\tabref{tab:coeffs_pos}. For $x=1/2$ we use the known value
$\li{3}(1/2)=[21\zeta(3)+4\ln^3(2)-2\pi^2\ln(2)]/24$. For
$x\in(1/2,1)$ we use Eq.~\eqref{eq:inversion2} to transform the
argument to the intervals $(-1,0)$ and $(0,1/2)$, where we use the
appropriate approximant from Eq.~\eqref{eq:minimax} on each interval.
For $x=1$ we use the known value $\li{3}(1)=\zeta(3)$.  For
$x\in(1,2)$ we use Eq.~\eqref{eq:inversion2} to transform the argument
to the intervals $(-1,0)$ and $(0,1/2)$, where we use the appropriate
approximant from Eq.~\eqref{eq:minimax} on each interval.  For
$x\geq 2$ we use Eq.~\eqref{eq:inversion1} to transform the argument
to the interval $(0,1/2]$, where we use the appropriate approximant
from Eq.~\eqref{eq:minimax}.
\begin{table}[t]
  \centering
  \caption{Coefficients of the numerator and denominator polynomials
    for the minimax approximant in Eq.~\eqref{eq:minimax} for
    $x\in[-1,0]$.}
  \begin{tabular}{lr}
    \toprule
    $p_0$ & $ 0.9999999999999999795\cdot 10^{+0}$ \\
    $p_1$ & $-2.0281801754117129576\cdot 10^{+0}$ \\
    $p_2$ & $ 1.4364029887561718540\cdot 10^{+0}$ \\
    $p_3$ & $-4.2240680435713030268\cdot 10^{-1}$ \\
    $p_4$ & $ 4.7296746450884096877\cdot 10^{-2}$ \\
    $p_5$ & $-1.3453536579918419568\cdot 10^{-3}$ \\
    $q_0$ & $ 1.0000000000000000000\cdot 10^{+0}$ \\
    $q_1$ & $-2.1531801754117049035\cdot 10^{+0}$ \\
    $q_2$ & $ 1.6685134736461140517\cdot 10^{+0}$ \\
    $q_3$ & $-5.6684857464584544310\cdot 10^{-1}$ \\
    $q_4$ & $ 8.1999463370623961084\cdot 10^{-2}$ \\
    $q_5$ & $-4.0756048502924149389\cdot 10^{-3}$ \\
    $q_6$ & $ 3.4316398489103212699\cdot 10^{-5}$ \\
    \bottomrule
  \end{tabular}
  \label{tab:coeffs_neg}
\end{table}
\begin{table}[t]
  \centering
  \caption{Coefficients of the numerator and denominator polynomials
    for the minimax approximant in Eq.~\eqref{eq:minimax} for
    $x\in[0,1/2]$.}
  \begin{tabular}{lr}
    \toprule
    $p_0$ & $ 0.9999999999999999893\cdot 10^{+0}$ \\
    $p_1$ & $-2.5224717303769789628\cdot 10^{+0}$ \\
    $p_2$ & $ 2.3204919140887894133\cdot 10^{+0}$ \\
    $p_3$ & $-9.3980973288965037869\cdot 10^{-1}$ \\
    $p_4$ & $ 1.5728950200990509052\cdot 10^{-1}$ \\
    $p_5$ & $-7.5485193983677071129\cdot 10^{-3}$ \\
    $q_0$ & $ 1.0000000000000000000\cdot 10^{+0}$ \\
    $q_1$ & $-2.6474717303769836244\cdot 10^{+0}$ \\
    $q_2$ & $ 2.6143888433492184741\cdot 10^{+0}$ \\
    $q_3$ & $-1.1841788297857667038\cdot 10^{+0}$ \\
    $q_4$ & $ 2.4184938524793651120\cdot 10^{-1}$ \\
    $q_5$ & $-1.8220900115898156346\cdot 10^{-2}$ \\
    $q_6$ & $ 2.4927971540017376759\cdot 10^{-4}$ \\
    \bottomrule
  \end{tabular}
  \label{tab:coeffs_pos}
\end{table}

An implementation of the described algorithm in C is given in
\appref{app:implementation}. This C implementation is also
provided in an ancillary file in the arXiv submission of this
publication under the CC-BY-4.0 license.

\section{Benchmark}

In the following we investigate the run\-/time of the C implementation
given in \appref{app:implementation}. \tabref{tab:runtime} shows the
average run\-/time of a single call of $\li{3}$ in nano seconds for
arguments on different intervals and on different 64-bit CPU
architectures (compiled with \texttt{gcc} 10.2.1 with \texttt{-O2}
optimization level). The run\-/times shown in the table have been
obtained by measuring the average run\-/time of $\li{3}$ on $10^6$
random values on each interval using the Google benchmark library
version 1.5.2 \cite{googlebenchmark}. For comparison we also show in
\tabref{tab:runtime} the average run\-/time for the real natural
logarithm $\ln$, the real cosine function $\cos$ and the real identity
function $\id(x)=x$.
\begin{table}[tb]
  \centering
  \caption{Average run\-/time in nano seconds for one invocation of
    $\li{3}$, $\ln$, $\cos$ and $\id$ for arguments on different
    intervals on different CPU architectures (compiled with
    \texttt{gcc} 10.2.1 with \texttt{-O2} optimization level).}
  \begin{tabular}{cccc}
    \toprule
    Function & Interval    & i5-8265U & i7-5600U \\
    \midrule
    $\li{3}$ & $[-2,-1]$   & $16.7$   & $20.3$   \\
    $\li{3}$ & $[-1,0]$    & $4.31$   & $5.81$   \\
    $\li{3}$ & $[0,1/2]$   & $5.09$   & $6.42$   \\
    $\li{3}$ & $[1/2,1]$   & $24.3$   & $29.4$   \\
    $\li{3}$ & $[1,2]$     & $24.2$   & $31.0$   \\
    $\li{3}$ & $[2,3]$     & $13.2$   & $15.9$   \\
    $\ln$    & $[1,2]$     & $5.20$   & $6.34$   \\
    $\cos$   & $[0,\pi/2]$ & $11.0$   & $13.3$   \\
    $\id$    & $[-2,2]$    & $0.259$  & $0.315$  \\
    \bottomrule
  \end{tabular}
  \label{tab:runtime}
\end{table}%

We find that the run\-/time of $\li{3}$ is similar to the run\-/time
of $\ln$ for arguments on the intervals $[-1,0]$ and $[0,1/2]$, where
no transformation is performed and the rational minimax approximants
from Eq.~\eqref{eq:minimax} are used directly. For arguments
$x\in[-2,-1]$ or $x\in[2,3]$ the transformation onto $[-1,0]$ and
$[0,1/2]$, respectively, involves extra arithmetic floating\-/point
operations and one additional call of $\ln$, which leads to an
increased run\-/time of $\li{3}$ by approximately a factor $3$ in
total. For arguments $x\in[1/2,1]$ or $x\in[1,2]$ a transformation
onto both $[-1,0]$ and $[0,1/2]$ intervals is performed, where the
corresponding approximants from Eq.~\eqref{eq:minimax} are used,
respectively.  This transformation, which requires extra arithmetic
floating\-/point operations and two additional calls of $\ln$, and the
necessity to use two approximants leads to an increased run\-/time of
$\li{3}$ by approximately a factor $5$ on these intervals.\footnote{A
  potential performance optimization for arguments $x\in[1/2,1]$ or
  $x\in[1,2]$ could be to not perform a transformation onto other
  intervals, but instead directly use dedicated rational minimax
  approximants on these intervals, at the cost of an increased number
  of coefficients to be stored in the source code.}

\section{Summary}

We have presented an algorithm to approximate the real trilogarithm of
a real argument with IEEE 754\-/1985 double precision. The
approximation is structured to allow for the use of
instruction\-/level parallelism on appropriate CPUs.
A C implementation of the real trilogarithm, using the algorithm
presented in this publication, can be found in the appendix as well as
in an ancillary file in the arXiv submission of this publication under
the CC-BY-4.0 license.  Implementations in C, C++, Fortran, Julia and
Rust can be found in Refs.~\cite{polylogarithm,PolyLog.jl,polylog.rs}.

\newpage
\appendix

\section{Implementation of the real trilogarithm}
\label{app:implementation}

\begin{lstlisting}[language=C]
#include <math.h>

// Re[Li_3(x)] for x in [-1, 0]
static double li3_neg(double x)
{
   const double P[] = {
      0.9999999999999999795e+0,
     -2.0281801754117129576e+0,
      1.4364029887561718540e+0,
     -4.2240680435713030268e-1,
      4.7296746450884096877e-2,
     -1.3453536579918419568e-3
   };
   const double Q[] = {
      1.0000000000000000000e+0,
     -2.1531801754117049035e+0,
      1.6685134736461140517e+0,
     -5.6684857464584544310e-1,
      8.1999463370623961084e-2,
     -4.0756048502924149389e-3,
      3.4316398489103212699e-5
   };

   const double x2 = x*x;
   const double x4 = x2*x2;
   const double p = P[0] + x*P[1]
      + x2*(P[2] + x*P[3])
      + x4*(P[4] + x*P[5]);
   const double q = Q[0] + x*Q[1]
      + x2*(Q[2] + x*Q[3])
      + x4*(Q[4] + x*Q[5] + x2*Q[6]);

   return x*p/q;
}

// Re[Li_3(x)] for x in [0, 1/2]
static double li3_pos(double x)
{
   const double P[] = {
      0.9999999999999999893e+0,
     -2.5224717303769789628e+0,
      2.3204919140887894133e+0,
     -9.3980973288965037869e-1,
      1.5728950200990509052e-1,
     -7.5485193983677071129e-3
   };
   const double Q[] = {
      1.0000000000000000000e+0,
     -2.6474717303769836244e+0,
      2.6143888433492184741e+0,
     -1.1841788297857667038e+0,
      2.4184938524793651120e-1,
     -1.8220900115898156346e-2,
      2.4927971540017376759e-4
   };

   const double x2 = x*x;
   const double x4 = x2*x2;
   const double p = P[0] + x*P[1]
      + x2*(P[2] + x*P[3])
      + x4*(P[4] + x*P[5]);
   const double q = Q[0] + x*Q[1]
      + x2*(Q[2] + x*Q[3])
      + x4*(Q[4] + x*Q[5] + x2*Q[6]);

   return x*p/q;
}

// Re[Li_3(x)] for x in (-inf, +inf)
double li3(double x)
{
   const double zeta2 = 1.6449340668482264;
   const double zeta3 = 1.2020569031595943;

   if (x < -1) {
      const double l = log(-x);
      return li3_neg(1/x)
         - l*(zeta2 + 1.0/6*l*l);
   } else if (x == -1) {
      return -0.75*zeta3;
   } else if (x < 0) {
      return li3_neg(x);
   } else if (x == 0) {
      return 0;
   } else if (x < 0.5) {
      return li3_pos(x);
   } else if (x == 0.5) {
      return 0.53721319360804020;
   } else if (x < 1) {
      const double l = log(x);
      return -li3_neg(1 - 1/x)
         - li3_pos(1 - x) + zeta3
         + l*(zeta2
              + l*(-0.5*log(1 - x)
                   + 1.0/6*l));
   } else if (x == 1) {
      return zeta3;
   } else if (x < 2) {
      const double l = log(x);
      return -li3_neg(1 - x)
         - li3_pos(1 - 1/x) + zeta3
         + l*(zeta2
              + l*(-0.5*log(x - 1)
                   + 1.0/6*l));
   } else {
      const double l = log(x);
      return li3_pos(1/x)
         + l*(2*zeta2 - 1.0/6*l*l);
   }
}
\end{lstlisting}

\printbibliography

@book{lewin,
    author       = {{Leonard Lewin}},
    title        = {{Polylogarithms and Associated Functions}},
    isbn         = {0444005501},
    year         = {1981},
    publisher    = {Elsevier Norh Holland, Inc.}
}

@software{polylogarithm,
    author       = {{Alexander Voigt}},
    title        = {{Polylogarithm}},
    year         = {2023},
    version      = {6.14.0},
    url          = {https://github.com/Expander/polylogarithm},
    note         = {[License: MIT]}
}

@software{PolyLog.jl,
    author       = {{Alexander Voigt}},
    title        = {{PolyLog.jl}},
    year         = {2023},
    version      = {2.3.1},
    url          = {https://github.com/Expander/PolyLog.jl},
    note         = {[License: MIT]}
}

@software{polylog.rs,
    author       = {{Alexander Voigt}},
    title        = {{polylog.rs}},
    year         = {2023},
    version      = {2.5.1},
    url          = {https://github.com/Expander/polylog.rs},
    note         = {[License: LGPL-3.0]}
}

@software{mathematica,
    author       = {{Wolfram Research{,} Inc.}},
    title        = {Mathematica},
    year         = {2021},
    version      = {13.0.0},
    note         = {Champaign, IL}
}

@software{googlebenchmark,
    author       = {{Google LLC}},
    title        = {{Benchmark}},
    year         = {2020},
    version      = {1.5.2},
    url          = {https://github.com/google/benchmark},
    note         = {[License: Apache-2.0]}
}

@inproceedings{estrin,
    author       = {Estrin, Gerald},
    title        = {{Organization of Computer Systems: The Fixed plus Variable Structure Computer}},
    year         = {1960},
    isbn         = {9781450378697},
    publisher    = {Association for Computing Machinery},
    address      = {New York, NY, USA},
    % url          = {https://doi.org/10.1145/1460361.1460365},
    doi          = {10.1145/1460361.1460365},
    abstract     = {The past decade has seen the development of productive fast electronic digital computers. Significant problems have been solved and significant numerical experiments have been executed. Moreover, as expected, a growing number of important problems have been recorded which are not practicably computable by existing systems. These latter problems have provided the incentive for the present development of several large scale digital computers with the goal of one or two orders of magnitude increase in overall computational speed.},
    booktitle    = {Papers Presented at the May 3-5, 1960, Western Joint IRE-AIEE-ACM Computer Conference},
    pages        = {33--40},
    numpages     = {8},
    location     = {San Francisco, California},
    series       = {IRE-AIEE-ACM '60 (Western)}
}

@book{luke,
    author       = {Yudell L. Luke},
    title        = {{Mathematical Functions and their Approximations}},
    doi          = {10.1016/C2013-0-11106-3},
    publisher    = {Academic Press Inc.},
    year         = {1975}
}

@article{clenshaw,
    title        = {{A note on the summation of Chebyshev series}},
    author       = {C. W. Clenshaw},
    journal      = {Mathematics of Computation},
    year         = {1955},
    volume       = {9},
    pages        = {118--120}
}

@article{morris,
    title        = {{The dilogarithm function of a real argument}},
    author       = {Robert A. Morris},
    journal      = {Mathematics of Computation},
    year         = {1979},
    volume       = {33},
    pages        = {778-787}
}

@article{Voigt:2022xnc,
    author       = {Alexander Voigt},
    title        = {{Comparison of methods for the calculation of the real dilogarithm regarding instruction-level parallelism}},
    doi          = {10.48550/arXiv.2201.01678},
    journal      = {arXiv},
    eprint       = {2201.01678},
    archivePrefix = {arXiv},
    primaryClass = {hep-ph},
    month        = {1},
    year         = {2022}
}

@article{Martin:2005qm,
    author       = {Martin, Stephen P. and Robertson, David G.},
    title        = {{TSIL: A Program for the calculation of two-loop self-energy integrals}},
    eprint       = {hep-ph/0501132},
    archivePrefix = {arXiv},
    reportNumber = {FERMILAB-PUB-05-683-T},
    doi          = {10.1016/j.cpc.2005.08.005},
    journal      = {Comput. Phys. Commun.},
    volume       = {174},
    pages        = {133--151},
    year         = {2006}
}

@article{Martin:2016bgz,
    author       = {Martin, Stephen P. and Robertson, David G.},
    title        = {{Evaluation of the general 3-loop vacuum Feynman integral}},
    eprint       = {1610.07720},
    archivePrefix = {arXiv},
    primaryClass = {hep-ph},
    doi          = {10.1103/PhysRevD.95.016008},
    journal      = {Phys. Rev. D},
    volume       = {95},
    number       = {1},
    pages        = {016008},
    year         = {2017}
}

@article{Buehler:2011ev,
    author       = {Buehler, Stephan and Duhr, Claude},
    title        = {{CHAPLIN - Complex Harmonic Polylogarithms in Fortran}},
    eprint       = {1106.5739},
    archivePrefix = {arXiv},
    primaryClass = {hep-ph},
    reportNumber = {IPPP-11-36, DCPT-11-72},
    doi          = {10.1016/j.cpc.2014.05.022},
    journal      = {Comput. Phys. Commun.},
    volume       = {185},
    pages        = {2703--2713},
    year         = {2014}
}

@article{Naterop:2019xaf,
    author       = {Naterop, L. and Signer, A. and Ulrich, Y.},
    title        = {{handyG -- Rapid numerical evaluation of generalised polylogarithms in Fortran}},
    eprint       = {1909.01656},
    archivePrefix = {arXiv},
    primaryClass = {hep-ph},
    reportNumber = {PSI-PR-19-17, ZU-TH 40/19},
    doi          = {10.1016/j.cpc.2020.107165},
    journal      = {Comput. Phys. Commun.},
    volume       = {253},
    pages        = {107165},
    year         = {2020}
}

@article{Ablinger:2018sat,
    author       = {Ablinger, J. and Bl\"umlein, J. and Round, M. and Schneider, C.},
    title        = {{Numerical Implementation of Harmonic Polylogarithms to Weight w = 8}},
    eprint       = {1809.07084},
    archivePrefix = {arXiv},
    primaryClass = {hep-ph},
    reportNumber = {DESY 13--064, DO--TH 17/12, DESY-13-064, DO-TH-17-12},
    doi          = {10.1016/j.cpc.2019.02.005},
    journal      = {Comput. Phys. Commun.},
    volume       = {240},
    pages        = {189--201},
    year         = {2019}
}

\end{document}